\documentclass[aps,twocolumn,amsmath,amssymb,a4paper]{revtex4}

\usepackage[utf8]{inputenc}
\usepackage[english]{babel}
\usepackage{amsmath}
\usepackage{amsfonts}
\usepackage{amssymb}
\usepackage{amsthm}
\usepackage{graphicx}
\usepackage{color}
\usepackage{xcolor}
\usepackage{fullpage}
\usepackage{fancybox}
\usepackage{float}
\usepackage{placeins}
\usepackage{tikz}
\usetikzlibrary{trees,arrows,decorations.pathmorphing,positioning,shapes}
\usetikzlibrary{decorations.markings,fadings}
\usepackage{verbatim}

\theoremstyle{definition}

\newcommand{\mr}{\mathbf{r}}

\newcommand{\wsr}{w_\text{ee}^{\sr}}
\newcommand{\bra}[1]{\ensuremath{\langle #1 \vert}}
\newcommand{\ket}[1]{\ensuremath{\vert #1  \rangle}}
\renewcommand{\H}{\ensuremath{\text{H}}}
\renewcommand{\b}[1]{\ensuremath{\mathbf{#1}}}
\newcommand{\KS}{\ensuremath{\text{KS}}}

\newcommand{\lr}{\ensuremath{\text{lr}}}
\newcommand{\sr}{\ensuremath{\text{sr}}}
\newcommand{\w}{w_\text{ee}}
\DeclareMathOperator{\erf}{erf}
\DeclareMathOperator{\erfc}{erfc}

\begin{document}

\title{Excitation energies along a range-separated adiabatic connection}
\author{Elisa Rebolini$^{1,2}$}\email{rebolini@lct.jussieu.fr}
\author{Julien Toulouse$^{1,2}$}\email{julien.toulouse@upmc.fr}
\author{Andrew M. Teale$^{3,4}$}
\author{Trygve Helgaker$^{4}$}
\author{Andreas Savin$^{1,2}$}\email{savin@lct.jussieu.fr}
\affiliation{
$^1$Sorbonne Universit\'es, UPMC Univ Paris 06, UMR 7616, Laboratoire de Chimie Th\'eorique, F-75005 Paris, France\\
$^2$CNRS, UMR 7616, Laboratoire de Chimie Th\'eorique, F-75005 Paris, France\\
$^3$School of Chemistry, University of Nottingham, University Park, Nottingham NG7 2RD, United Kingdom\\
$^4$Centre for Theoretical and Computational Chemistry, Department of Chemistry, University of Oslo, P.O. Box 1033 Blindern, N-0315 Oslo, Norway}

\date{July 9, 2014}

\begin{abstract}
We present a study of the variation of total energies and excitation
energies along a range-separated adiabatic connection. This connection
links the non-interacting Kohn--Sham electronic system to the physical
interacting system by progressively switching on the
electron--electron interactions whilst simultaneously adjusting a
one-electron effective potential so as to keep the ground-state
density constant. The interactions are introduced in a
range-dependent manner, first introducing predominantly long-range,
and then all-range, interactions as the physical system is approached,
as opposed to the conventional adiabatic connection where the
interactions are introduced by globally scaling the standard Coulomb interaction.
Reference data are reported for the He and Be atoms and the H$_2$
molecule, obtained by calculating the short-range effective potential
at the full configuration-interaction level using Lieb's
Legendre-transform approach. As the strength of the electron--electron
interactions increases, the excitation energies, calculated for the
partially interacting systems along the adiabatic connection, offer
increasingly accurate approximations to the exact excitation energies.
Importantly, the excitation energies calculated at an intermediate
point of the adiabatic connection are much better approximations to
the exact excitation energies than are the corresponding Kohn--Sham
excitation energies. This is particularly evident in situations
involving strong static correlation effects and states with multiple
excitation character, such as the dissociating H$_2$ molecule. These
results highlight the utility of long-range interacting reference
systems as a starting point for the calculation of excitation energies
and are of interest for developing and analyzing practical approximate
range-separated density-functional methodologies.
\end{abstract}

\maketitle

\section{Introduction}

Range-separated density-functional theory (see, e.g.,
Ref.\;\onlinecite{TouColSav-PRA-04}) constitutes an interesting
alternative to standard Kohn--Sham (KS) density-functional theory
(DFT)~\cite{HohKoh-PR-64,KohSha-PR-65}. In the standard KS approach,
the physical interacting electronic Hamiltonian is replaced by an
effective non-interacting Hamiltonian. By contrast, in range-separated
DFT, the physical Hamiltonian is instead replaced by a partially
interacting Hamiltonian that incorporates the long-range part of the
electron--electron interaction. This corresponds to an intermediate
point along a range-separated adiabatic
connection~\cite{Sav-INC-96,Yan-JCP-98,PolColLeiStoWerSav-IJQC-03,SavColPol-IJQC-03,TouColSav-PRA-04}.
The KS Hamiltonian is linked to the physical Hamiltonian by
progressively switching on the long-range part of the two-electron
interaction, whilst simultaneously modifying the one-electron
potential so as to maintain a constant ground-state density.  The
ground-state energy of the physical system can then be extracted from
the ground state of the long-range interacting Hamiltonian by using a
short-range density functional describing the complementary
short-range part of the electron--electron interaction. Be aware that this
range-separated manner of introducing the interaction is not the usual
way of performing the adiabatic connection, where the Coulomb interaction is
instead scaled by a multiplicative constant going from 0 to 1.

Several short-range density-functional approximations have been
developed~\cite{Sav-INC-96,TouSavFla-IJQC-04,TouColSav-PRA-04,TouColSav-JCP-05,TouGorSav-TCA-05,GolWerSto-PCCP-05,PazMorGorBac-PRB-06,GolErnMoeSto-JCP-09}
and a diverse range of approaches for calculating the ground state of
the long-range interacting Hamiltonian have been explored. To aid in
the description of static (or strong) correlation effects, which are
poorly treated by standard density functionals,
configuration-interaction~\cite{SavFla-IJQC-95,Sav-INC-96a,Sav-INC-96,LeiStoWerSav-CPL-97,PolSavLeiSto-JCP-02,SavColPol-IJQC-03,TouColSav-PRA-04},
multiconfiguration self-consistent-field
(MCSCF)~\cite{FroTouJen-JCP-07,FroReaWahWahJen-JCP-09,StoTeaTouHelFro-JCP-13},
density-matrix functional theory
(DMFT)~\cite{Per-PRA-10,RohTouPer-PRA-10,RohPer-JCP-11}, and
constrained-pairing mean-field
theory~\cite{TsuScuSav-JCP-10,TsuScu-JCP-11} descriptions of the
long-range interacting systems have been employed. To treat van der
Waals interactions, second-order perturbation
theory~\cite{AngGerSavTou-PRA-05,GerAng-CPL-05b,GerAng-JCP-07,Ang-PRA-08,FroJen-PRA-08,GolLeiManMitWerSto-PCCP-08,JanScu-PCCP-09,FroCimJen-PRA-10,ChaStoWerLei-MP-10,ChaJacAdaStoLei-JCP-10,FroJen-JCP-11,CorStoJenFro-PRA-13},
coupled-cluster
theory~\cite{GolWerSto-PCCP-05,GolWerStoLeiGorSav-CP-06,GolStoThiSch-PRA-07,GolWerSto-CP-08,GolErnMoeSto-JCP-09},
and random-phase
approximations~\cite{TouGerJanSavAng-PRL-09,JanHenScu-JCP-09,JanHenScu-JCP-09b,JanScu-JCP-09,ZhuTouSavAng-JCP-10,TouZhuAngSav-PRA-10,PaiJanHenScuGruKre-JCP-10,TouZhuSavJanAng-JCP-11,AngLiuTouJan-JCTC-11,IreHenScu-JCP-11,GouDob-PRB-11}
have been used successfully.

Electronic excitation energies can also be calculated in
range-separated DFT by using the linear-response approach with a
time-dependent generalization of the static ground-state
theory~\cite{FroKneJen-JCP-13}. In this case, the excitation energies
of the long-range interacting Hamiltonian act as starting
approximations that are then corrected using a short-range
density-functional kernel, just as the KS excitation energies act as
starting approximations in linear-response time-dependent
density-functional theory (TDDFT). Several such range-separated
linear-response schemes have been developed, in which the short-range
part is described by an approximate adiabatic semi-local
density-functional kernel and the long-range linear-response part is
treated at the
Hartree--Fock~\cite{RebSavTou-MP-13,TouRebGouDobSeaAng-JCP-13,FroKneJen-JCP-13,HedHeiKneFroJen-JCP-13},
MCSCF~\cite{FroKneJen-JCP-13,HedHeiKneFroJen-JCP-13}, second-order
polarization-propagator approximation
(SOPPA)~\cite{HedHeiKneFroJen-JCP-13}, or DMFT~\cite{Per-JCP-12}
levels. These schemes aim at overcoming the limitations of standard
linear-response TDDFT applied with usual adiabatic semi-local
approximations for describing systems with static
correlation~\cite{GriGisGorBae-JCP-00}, double or multiple
excitations~\cite{MaiZhaCavBur-JCP-04}, and Rydberg or charge-transfer
excitations~\cite{CasJamCasSal-JCP-98,DreWeiHea-JCP-03}.

For the purpose of analyzing linear-response range-separated DFT
approaches, it is desirable to have accurate reference values of the
excitation energies of the long-range interacting Hamiltonian along
the range-separated adiabatic connection [cf.~Eq.\;(\ref{Hmu})]. In
this work, we provide and analyze reference data for the He and Be
atoms and the H$_2$ molecule. The short-range one-electron potentials
required to keep the ground-density constant along a range-separated
adiabatic connection [cf.~Eq.\;(\ref{VHxcsrmu})] are determined at the
full configuration-interaction (FCI) level using Lieb's
Legendre-transform
approach~\cite{Lie-IJQC-83,ColSav-JCP-99,TeaCorHel-JCP-09}. The
excited-state energies of the long-range interacting Hamiltonian along
the adiabatic connection [cf.~Eq.\;(\ref{HlrmuPsi})] are then
calculated using the FCI method. Several accurate ground-state
calculations have been performed in the past along the standard
adiabatic
connection~\cite{JouSri-JCP-98,ColSav-JCP-99,SavColAll-JCP-01,TeaCorHel-JCP-09,TeaCorHel-JCP-10,StrKumCorSagTeaHel-JCP-11}
and range-separated adiabatic
connections~\cite{PolColLeiStoWerSav-IJQC-03,TouColSav-PRA-04,TouColSav-MP-05,TeaCorHel-JCP-10b,StrKumCorSagTeaHel-JCP-11}
for small atomic and molecular systems, but accurate calculations of
excited-state energies along adiabatic connections are very
scarce---see, however, Refs.~\onlinecite{ColSav-JCP-99,ZhaBur-PRA-04}.

The remainder of this paper is organized as follows. In
Section~\ref{sec:rsdft}, range-separated DFT is briefly reviewed and
the definition of the excited states along the range-separated
adiabatic connection is introduced. In Section~\ref{app:smalllargemu},
the behaviour of the excited-state energies near the two endpoints of
the adiabatic connection, the KS system and the physical system, is
studied analytically. After giving computational details in
Section~\ref{sec:computational}, results along the full
adiabatic-connection path are presented and discussed in
Section~\ref{sec:results}. Finally, some concluding remarks are made
in Section~\ref{sec:conclusion}.

\section{Range-separated density-functional theory}
\label{sec:rsdft}

In range-separated DFT (see, e.g.,
Ref.~\onlinecite{TouColSav-PRA-04}), the exact ground-state energy of
an $N$-electron system is obtained by the following minimization over
normalized multi-determinantal wave functions $\Psi$:
\begin{eqnarray}
E_0 &=& \min_{\Psi} \Bigl\{ \bra{\Psi} \hat{T} + \hat{V}_\text{ne} +
\hat{W}_\text{ee}^{\lr,\mu} \ket{\Psi} + \bar{E}_{\H
  \text{xc}}^{\sr,\mu}[n_{\Psi}]\Bigl\}.  \nonumber\\
\label{EminPsi}
\end{eqnarray}
This expression contains the kinetic-energy operator $\hat{T}$, the
nuclear--electron interaction operator $\hat{V}_\text{ne} = \int
v_\text{ne}(\b{r}) \hat{n}(\b{r}) \mathrm d\b{r}$ expressed in terms
of the density operator $\hat{n}(\b{r})$, and a long-range (lr)
electron--electron interaction operator
\begin{eqnarray}
\hat{W}_\text{ee}^{\lr,\mu} \!=\! \frac{1}{2} \iint
w_\text{ee}^{\lr,\mu}(r_{12}) \hat{n}_2(\b{r}_1,\b{r}_2) \mathrm
d\b{r}_1 \mathrm d\b{r}_2,
\end{eqnarray}
expressed in terms of the pair-density operator $\hat{n}_2(\b{r}_1,\b{r}_2)$. 
In the present work, we use the error-function interaction
\begin{equation}
w_\text{ee}^{\lr,\mu}(r_{12}) = \frac{\erf(\mu r_{12})}{r_{12}},
\end{equation}
where $\mu$ controls the range of the separation, with $1/\mu$ acting
as a smooth cut-off radius. The corresponding complementary
short-range (sr) Hartree--exchange--correlation density functional
$\bar{E}_{\H \text{xc}}^{\sr,\mu}[n_\Psi]$ is evaluated at the density
of $\Psi$: $n_\Psi(\b{r}) = \bra{\Psi} \hat{n}(\b{r}) \ket{\Psi}$.

The Euler--Lagrange equation for the minimization of
Eq.~(\ref{EminPsi}) leads to the (self-consistent) eigenvalue equation
\begin{equation}
  \hat{H}^{\lr,\mu} | \Psi_0^{\mu} \rangle = {\cal E}_0^{\mu} | \Psi_0^{\mu} \rangle,
\end{equation}
where $\Psi_0^{\mu}$ and ${\cal E}_0^{\mu}$ are the ground-state wave
function and associated energy of the partially interacting
Hamiltonian (with an explicit long-range electron--electron
interaction)
\begin{equation}
  \hat{H}^{\lr,\mu} = \hat{T} + \hat{V}_\text{ne} + \hat{W}^{\lr,\mu}_\text{ee} + \hat{\bar{V}}_{\H \text{xc}}^{\sr,\mu}.
\label{Hmu}
\end{equation}
It contains the short-range Hartree--exchange--correlation potential operator, evaluated at the density $n_0(\b{r}) = \bra{\Psi_0^{\mu}} \hat{n}(\b{r}) \ket{\Psi_0^{\mu}}$, which is equal to the ground-state density of the physical system for all $\mu$,
\begin{equation}
\hat{\bar{V}}_{\H \text{xc}}^{\sr,\mu}= \int \bar{v}^{\sr,\mu}_{\H \text{xc}}[n_0](\mr) \hat{n}(\mr) \mathrm d\mr,
\label{VHxcsrmu}
\end{equation}
where 
\begin{equation}
\bar{v}^{\sr,\mu}_{\H \text{xc}}[n](\mr)  = \frac{\delta \bar{E}_{\H \text{xc}}^{\sr,\mu}[n]}{ \delta n(\mr)}.
\end{equation}
For $\mu=0$, $\hat{H}^{\lr,\mu}$ reduces to the standard non-interacting KS Hamiltonian, 
$\hat{H}^{\KS}$, while for $\mu\to\infty$ it reduces to the physical Hamiltonian $\hat{H}$:
\begin{alignat}{2}
\hat{H}^{\KS} &= \hat{H}^{\lr,\mu=0} &&= \hat{T} + \hat{V}_\text{ne} + \hat{V}_{\H \text{xc}} ,\\
\hat{H} &= \hat{H}^{\lr,\mu= \infty} &&= \hat{T} + \hat{V}_\text{ne} + \hat{W}_{\text{ee}} .
\label{KSham}
\end{alignat}
Varying the parameter $\mu$ between these two limits, $\hat{H}^{\lr,\mu}$ defines a range-separated adiabatic connection, linking the non-interacting KS system to the physical 
system with the ground-state density kept constant (provided that the exact short-range Hartree--exchange--correlation potential $\bar{v}_{\H \text{xc}}^{\sr,\mu}(\mr)$ is used). 

In this work we also consider the excited-state wave functions and energies of the long-range interacting Hamiltonian
\begin{equation}
\hat{H}^{\lr,\mu} | \Psi_k^{\mu} \rangle = {\cal E}_k^{\mu} | \Psi_k^{\mu} \rangle,
\label{HlrmuPsi}
\end{equation}
where $\hat{H}^{\lr,\mu}$ is Hamiltonian in Eq.~(\ref{Hmu}), with the short-range Hartree--exchange--correlation potential evaluated at the \emph{ground-state density} $n_0$. 
In range-separated DFT, these excited-state wave functions and energies provide a natural first approximation to the excited-state wave functions and energies of the physical system. For $\mu=0$, they reduce to the single-determinant eigenstates and associated energies of the non-interacting KS Hamiltonian,
\begin{equation}
\hat{H}^{\KS} | \Phi_k^{\KS} \rangle = {\cal E}_k^{\KS} | \Phi_k^{\KS} \rangle,
\end{equation}
while, for $\mu \to \infty$, they reduce to the excited-state wave functions and energies of the physical Hamiltonian
\begin{equation}
\hat{H} | \Psi_k \rangle = E_k | \Psi_k \rangle.
\end{equation}
We emphasize that, even with the exact (short-range)
Hartree-exchange-correlation potential, the total energies ${\cal
  E}_k^{\KS}$ (${\cal E}_k^\mu$) are not the exact energies of the
physical system but the total energies of a non-interacting (partially
interacting) fictitious system of electrons with Hamiltonian $\hat{
  H}^\KS$ ($\hat{H}^{\lr,\mu}$).  Note also that, since the ionization
energy is related to the asymptotic decay of the ground-state density,
the ionization energy of the Hamiltonian in Eq.~(\ref{HlrmuPsi}) is
independent of $\mu$ and is equal to the ionization energy of the
physical system. This is an appealing feature since it sets the
correct energy window for bound excited states. Finally, note that the
excitation energies $\Delta {\cal E}_k^{\mu} = {\cal E}_k^{\mu} -
{\cal E}_0^{\mu}$ calculated from Eq.~(\ref{HlrmuPsi}) constitute a
starting point for range-separated linear-response theory based on the
time-dependent generalization of
Eq.~(\ref{EminPsi})~\cite{FroKneJen-JCP-13}.

\section{Excited-state energies near the Kohn--Sham and physical systems}
\label{app:smalllargemu}

In this section, we study analytically the behavior of the excited-state energies ${\cal E}_k^{\mu}$ as a function of the range-separation parameter $\mu$ close to the endpoints of the adiabatic connection: the KS system at $\mu=0$ and the physical system at $\mu \to \infty$. This study will aid in the understanding of the numerical results presented in Section~\ref{sec:results}.

\subsection{Excited-state energies near the Kohn--Sham system} 

We first derive the expansion of the excited-state energies near $\mu=0$, to see how the KS energies are affected by the introduction 
of the long-range electron--electron interaction. We assume that the system is spatially finite.

We rewrite the long-range interacting Hamiltonian of Eq.~(\ref{Hmu}) as
\begin{equation}
\hat{H}^{\lr,\mu} = \hat{H}^{\KS} + \hat{W}_\text{ee}^{\lr,\mu} - \hat{V}_{\H \text{xc}}^{\lr,\mu},
\label{Hlrmu0}
\end{equation}
with the long-range Hartree--exchange--correlation potential operator
\begin{equation}
\hat{V}_{\H \text{xc}}^{\lr,\mu} = \hat{V}_{\H \text{xc}} - \hat{\bar{V}}_{\H
  \text{xc}}^{\sr,\mu} = \int \!\!v_{\H \text{xc}}^{\lr,\mu} (\b{r}) \hat{n}(\b{r})
\mathrm d\b{r} .
\end{equation}
The expansion of the long-range two-electron interaction is
straightforward~\cite{TouColSav-PRA-04} (valid for $\mu r_{12} \ll 1$)
\begin{align}
  w_\text{ee}^{\lr,\mu}(r_{12}) &= \frac{\erf(\mu r_{12})}{r_{12}} \nonumber \\
   &= \frac{2\mu}{\sqrt{\pi}} + \mu^3 w_\text{ee}^{\lr,(3)}(r_{12}) + {\cal O} (\mu^5),
\label{weelrexpand}
\end{align}
with
\begin{align} 
w_\text{ee}^{\lr,(3)}(r_{12})&=-\frac{2}{3\sqrt{\pi}} r_{12}^2 .
\end{align}
Note that the first term in the expansion of
$w_\text{ee}^{\lr,\mu}(r_{12})$ in Eq.~(\ref{weelrexpand}) is a
spatial constant, $2\mu/\sqrt{\pi}$, which shows that what we call the
long-range interaction does in fact contain also a contribution at
short range~\cite{TouColSav-PRA-04}.  Next, the expansion of the
long-range Hartree--exchange--correlation potential
\begin{equation}
v_{\H \text{xc}}^{\lr,\mu}(\b{r}) =
\frac{\delta E_{\H \text{xc}}^{\lr,\mu}[n]}{\delta n(\b{r})}
\end{equation}
can be determined from the
expansion of the corresponding energy
functional $E_{\H \text{xc}}^{\lr,\mu}[n]$. As derived in
Ref.~\onlinecite{TouColSav-PRA-04}, the expansion of the 
Hartree--exchange part begins at first order and may be written as
\begin{eqnarray}
    E_{\H \text x}^{\lr,\mu}[n] &=& \dfrac{\mu}{\sqrt{\pi}} \iint
    n_2^{\KS}(\mr_1,\mr_2) \mathrm d \mr_1 \mathrm d \mr_2
\nonumber\\
 &&+ \dfrac{\mu^{3}}{2} \iint n_2^{\KS}(\mr_1,\mr_2) w_\text{ee}^{\lr,(3)}(r_{12}) \mathrm d \mr_1 \mathrm d \mr_2
\nonumber\\
&& + \mathcal{O}(\mu^5).
     \label{EHxlrmu}
\end{eqnarray}
where $n_{2}^{\KS}(\mr_1,\mr_2)$ is the KS pair density, while the
expansion of the correlation part only begins at sixth order (assuming a non-degenerate KS ground state)
\begin{eqnarray}
  E_\text c^{\lr,\mu}[n]= 0 + \mathcal{O}(\mu^6). 
  \label{Eclrmu}
\end{eqnarray}
If the functional derivative of $E_{\H \text x}^{\lr,\mu}[n]$
is taken with respect to density variations that preserve the number of electrons, $\int \!\delta n(\b{r}) \mathrm d\b{r}=0$, 
then the first-order term in Eq.\;(\ref{EHxlrmu}) does not contribute due to the fixed normalization of the KS pair density, $\iint n_2^{\KS}(\mr_1,\mr_2) \mathrm d \mr_1 \mathrm d \mr_2 = N(N-1)$. The derivative is then defined up to an additive ($\mu$-dependent) constant $C^\mu$, which can be fixed by requiring that a distant electron experiences zero potential interaction in Eq.~(\ref{Hlrmu0}), amounting to setting the zero-energy reference. The linear term in $\mu$ in the long-range Hartree--exchange--correlation potential can then be determined as follows. 

To first order in $\mu$, the long-range electron--electron interaction tends to a constant, $2\mu/\sqrt{\pi}$.
A distant electron (with $1 \ll r_{12}\ll 1/\mu$) then experiences a constant interaction $2(N-1) \mu/\sqrt{\pi}$ 
with the remaining $N-1$ other electrons. This constant must be exactly compensated 
by the long-range Hartree--exchange--correlation potential in Eq.~(\ref{Hlrmu0}),
so that its first-order term in $\mu$ must also be $2(N-1) \mu/\sqrt{\pi}$. The expansion of $v_{\H x c}^{\lr,\mu}(\mr)$ therefore takes the form
\begin{eqnarray}
    v_{\H \text{xc}}^{\lr,\mu}(\mr) &=& \dfrac{2(N-1) \mu}{\sqrt{\pi}} + \mu^3 v_{\H \text{xc}}^{\lr,(3)}(\b{r}) + \mathcal{O}(\mu^5),
\nonumber\\
\label{eq:vHx0}
\end{eqnarray}
where $v_{\H \text{xc}}^{\lr,(3)}(\b{r})$ is the third-order contribution.

Combining Eqs.\;(\ref{weelrexpand}) and (\ref{eq:vHx0}), we arrive at the
following expansion of the long-range interacting Hamiltonian of Eq.~(\ref{Hlrmu0}):
\begin{equation}
  \hat{H}^{\lr,\mu} = \hat{H}^{\KS} + \mu \hat{H}^{\lr,(1)} + \mu^3
  \hat{H}^{\lr,(3)} + \mathcal{O}(\mu^5),
\end{equation}
with a constant first-order correction
\begin{equation}
\hat{H}^{\lr,(1)} = -\frac{N(N-1)}{\sqrt{\pi}}
\end{equation}
and the following third-order correction
\begin{align}
\hat{H}^{\lr,(3)} &= \hat{W}_\text{ee}^{\lr,(3)} - \hat{V}_{\H
  \text{xc}}^{\lr,(3)}, \\ 
  \hat{W}_\text{ee}^{\lr,(3)} &= \frac{1}{2} \!\iint \!\!w_\text{ee}^{\lr,(3)}(r_{12})
  \hat{n}_2(\mr_1,\mr_2) \mathrm d\mr_1 \mathrm d\mr_2, \\
  \hat{V}_{\H \text{xc}}^{\lr,(3)} &= \int \!v_{\H \text{xc}}^{\lr,(3)}(\b{r})
  \hat{n}(\b{r}) \mathrm d\b{r}.
\end{align}
Since the first-order correction in the Hamiltonian is a constant, it
does not affect the associated wave functions. The expansion of the wave functions
therefore begins at third order in $\mu$:
\begin{equation}
  \Psi_k^{\mu} = \Phi_k^{\KS} + \mu^3 \Psi_k^{(3)} +\mathcal{O}(\mu^5).
  \label{eq:TaylorPsi0}
\end{equation}
Using normalized KS wave functions $\langle \Phi_k^{\KS} | \Phi_k^{\KS} \rangle =1$, the expansion of the total energy for the state $k$ is then
\begin{eqnarray}
    \mathcal{E}_k^{\mu} & = &\mathcal{E}_k^{\KS} -
    \dfrac{N(N-1)}{\sqrt{\pi}} \mu \nonumber\\ && +\mu^3 \langle
    \Phi_k^{\KS} |\hat{H}^{\lr,(3)} |\Phi_k^{\KS} \rangle+
    \mathcal{O}(\mu^5).
    \label{eq:Ei(mu=0)}
\end{eqnarray}
The first-order contribution is the same for all states, cancelling out in the differences between the energies of two states. 
As a result, the corrections to the KS excitation energies are third order in $\mu$.

For closed shells, the expansion of the difference between the singlet and triplet energies associated with the single excitation $i \to a$ 
can be obtained by applying Eq.~(\ref{eq:Ei(mu=0)}) with the spin-adapted KS wave functions ${^1}\Phi^\KS = \left( \Phi^\KS_{i\to a} + \Phi^\KS_{\bar{i}\to \bar{a}} \right) /\sqrt{2}$, for the singlet state, and ${^{3,1}}\Phi^\KS = \Phi^\KS_{\bar{i}\to a}$, for the triplet state with spin projection $M_S=1$. Only the two-electron term then contributes:
\begin{eqnarray}
    \Delta {\cal E}_{i \to a}^{\mu,1-3}
    &=& 2\mu^3 \bra{ia}\hat{w}_\text{ee}^{\lr,(3)}\ket{ai} + \mathcal{O}(\mu^5) 
\nonumber\\
    &=& \dfrac{8\mu^3}{3\sqrt{\pi}} \left| \bra{i} \hat{\mr} \ket{a} \right|^2
    + \mathcal{O}(\mu^5),
    \label{eq:spin splitting}
\end{eqnarray}
where we have used $r_{12}^2 = r_1^2+r_2^2-2\mr_1\cdot\mr_2$. The appearance of the transition dipole moment integral in Eq.~(\ref{eq:spin splitting}) means that, for an atomic system, the singlet--triplet energy splitting appears at third order in $\mu$ if the difference between the angular moment of the orbitals $\varphi_i$ and $\varphi_a$ is  $\Delta \ell = +1$ or $-1$. Otherwise, the splitting appears at a higher order in~$\mu$.

\subsection{Excited-state energies near the physical system} 

We now derive the asymptotic expansion of the excited-state energies
when $\mu \to \infty$, which shows how the exact excited-state
energies are affected by the removal of the very short-range part of
the electron--electron interaction.

For this purpose, we rewrite the long-range interacting Hamiltonian of Eq.\;(\ref{Hmu}) as
\begin{equation}
\hat{H}^{\lr,\mu} = \hat{H} - \hat{W}_\text{ee}^{\sr,\mu} + \hat{\bar{V}}_{\H \text{xc}}^{\sr,\mu},
\label{Hlrmuinf}
\end{equation}
where $\hat{H}$ is the Hamiltonian of the physical system, 
\begin{equation}
\hat{W}_\text{ee}^{\sr,\mu} \!=\! \frac{1}{2} \iint w_\text{ee}^{\sr,\mu}(r_{12}) \hat{n}_2(\b{r}_1,\b{r}_2) \mathrm d\b{r}_1 \mathrm d\b{r}_2 
\end{equation}
is the short-range electron--electron interaction operator defined with the complementary error-function interaction 
\begin{equation}
w_\text{ee}^{\sr,\mu}(r_{12}) = \frac{\erfc(\mu r_{12})}{r_{12}},
\end{equation}
and $\hat{\bar{V}}_{\H \text{xc}}^{\sr,\mu}$ is the short-range Hartree--exchange--correlation potential operator in Eq.~(\ref{VHxcsrmu}). 
The first term in the asymptotic expansion of $w_\text{ee}^{\sr,\mu}(r_{12})$ can be written in terms of a delta function~\cite{TouColSav-PRA-04} (valid for $\mu r_{12} \gg 1$)
\begin{equation}
  \begin{split}
    w_\text{ee}^{\sr,\mu}(r_{12}) = &\dfrac{\pi}{\mu^2}\delta(\b{r}_{12}) 
    + \mathcal{O}\left(\dfrac{1}{\mu^3}\right),
  \end{split}
\label{weesrexpand}
\end{equation}
while the expansion of $\bar{v}_{\H \text{xc}}^{\sr,\mu}(\b{r}) =  \delta \bar{E}_{\H \text{xc}}^{\sr,\mu}[n] /\delta n(\b{r})$ 
can be obtained from that of $\bar{E}_{\H \text{xc}}^{\sr,\mu}[n]$. As derived in Ref.~\onlinecite{TouColSav-PRA-04}, the expansion of the long-range Hartree--exchange energy is
\begin{equation}
  \begin{split}
    E_{\H x}^{\sr,\mu}[n]
    & = \dfrac{\pi}{2\mu^2} \int \! n_2^{\KS}(\mr,\mr) \mathrm d\mr 
    + \mathcal{O}\left( \dfrac{1}{\mu^4}\right),
  \end{split}
\label{EHxsrmu}
\end{equation}
where $n_2^{\KS}(\mr,\mr)$ is the KS on-top pair density, while the expansion of the long-range correlation energy is
\begin{equation}
  \begin{split}
    \bar{E}_c^{\sr,\mu}[n] 
    & = \dfrac{\pi}{2\mu^2}\int \! n_{2,c}(\mr, \mr) \mathrm d\mr 
    + \mathcal{O}\left( \dfrac{1}{\mu^3} \right), 
  \end{split}
\label{Ecsrmu}
\end{equation}
where $n_{2,c}(\mr, \mr)$ is the on-top correlation pair density of the physical system. Therefore, the expansion of the short-range Hartree--exchange--correlation potential takes the form
\begin{eqnarray}
  \bar{v}_{\H \text{xc}}^{\sr,\mu}(\mr) = \frac{1}{\mu^2} \bar{v}_{\H \text{xc}}^{\sr,(-2)}(\mr) + \mathcal{O}\left( \dfrac{1}{\mu^3} \right),
\label{vHxcsrexpand}
\end{eqnarray}
where $\bar{v}_{\H \text{xc}}^{\sr,(-2)}(\mr)$ is the $\mu^{-2}$ contribution formally obtained by taking the functional derivative of Eqs.~(\ref{EHxsrmu}) and~(\ref{Ecsrmu}).

Substituting Eqs.~(\ref{weesrexpand}) and~(\ref{vHxcsrexpand}) into Eq.~(\ref{Hlrmuinf}), we obtain the asymptotic expansion of the long-range interacting Hamiltonian as
\begin{equation}
  \label{eq:hmu infty}
  \begin{split}
    \hat{H}^{\lr,\mu}  & = \hat{H} + \dfrac{1}{\mu^2} \hat{H}^{\lr,(-2)} + \mathcal{O}\left( \dfrac{1}{\mu^3} \right),
  \end{split}
\end{equation}
where $\hat{H}^{\lr,(-2)}= - \hat{W}_\text{ee}^{\sr,(-2)} + \hat{\bar{V}}_{\H \text{xc}}^{\sr,(-2)}$ is composed of an on-top two-electron term
and a one-electron term:
\begin{align}
    \hat{W}_\text{ee}^{\sr,(-2)} &= \frac{\pi}{2}\int \hat{n}_2(\mr,\mr) \mathrm d\mr, \\
  \hat{\bar{V}}_{\H \text{xc}}^{\sr,(-2)} &= \int \bar{v}_{\H \text{xc}}^{\sr,(-2)}(\mr) \hat{n}(\b{r}) \mathrm d\b{r}.
\end{align}
The expansion of the Hamiltonian in Eq.~(\ref{eq:hmu infty}) suggests a similar expansion for the excited-state 
wave functions, $\Psi_k^\mu = \Psi_k + \mu^{-2}\Psi_k^{(-2)} + \mathcal{O}(\mu^{-3})$. However, as shown in Ref.~\onlinecite{Gori-Giorgi2006}, this expansion is not valid for $r_{12} \ll 1/\mu$. The contribution of the wave function for small $r_{12}$ to the integral for the total energy ${\cal E}^\mu_k = \bra{\Psi_k^\mu} \hat{H}^{\lr,\mu} \ket{\Psi_k^\mu}$ nevertheless vanishes in the limit $\mu \to \infty$, and the asymptotic expansion of the total energy of the state $k$ is
\begin{equation}
  {\cal E}_k^{\mu}= E_k
  + \dfrac{1}{\mu^2} \bra{\Psi_k} \hat{H}^{\lr,(-2)} \ket{\Psi_k}
  + \mathcal{O}\left(\dfrac{1}{\mu^3} \right)
  \label{eq:E0infty},
\end{equation}
where the wave function $\Psi_k$ is normalized to unity.

\section{Computational details}
\label{sec:computational}

Calculations have been performed for the He and Be atoms and for the
H$_2$ molecule with a development version of the DALTON
program~\cite{Dalshort-PROG-13,Dalton-WIRES-13}, using the
implementation described in Refs.~\onlinecite{TeaCorHel-JCP-09}
and~\onlinecite{TeaCorHel-JCP-10b}.  First, a FCI calculation was
performed to determine the exact ground-state density within the basis
set considered, followed by a Lieb optimization~\cite{ColSav-JCP-99}
of the short-range potential $v^{\sr,\mu}(\b{r})=v_\text{ne}(\b{r}) +
\bar{v}_{\H \text{xc}}^{\sr,\mu}(\b{r})$ also at the FCI level to
reproduce the FCI ground-state density in the presence of the
long-range electron--electron interaction $\w^{\lr,\mu}(r_{12})$.  The
FCI excited-state energies were then calculated using the partially
interacting Hamiltonian with the interaction $\w^{\lr,\mu}(r_{12})$
and effective potential $v^{\sr,\mu}(\b{r})$.

The Lieb maximization was performed using the short-range analogue of
the algorithm of Wu and Yang~\cite{Wu2003}, in which the potential is
expanded as
\begin{equation}
  v^{\sr,\mu}(\mr) = v_{\text{ne}}(\mr)+v^{\sr,\mu}_{\text{ref}}(\mr)
  + \sum_t b_t g_t(\mr).
\end{equation}
where the reference potential is the short-range analogue of the
Fermi--Amaldi potential
\begin{equation}
  v^{\sr,\mu}_{\text{ref}}(\mr) = \dfrac{N-1}{N} \int \! n_0(\mr')
  \wsr(|\mr - \mr'|) \mathrm d\mr' ,
\end{equation}
calculated for a fixed $N$-electron density $n_0$, to ensure the
correct asymptotic behaviour. The same Gaussian basis set $\{g_t\}$ is
used for the expansion of the potential and the molecular
orbitals. The coefficients $b_t$ are optimized by the Newton method,
using a regularized Hessian with a truncated
singular-value-decomposition cutoff of $10^{-7}$ for He and $10^{-6}$
for Be and H$_2$.

Even-tempered Kaufmann basis sets~\cite{Kaufmann1991} and uncontracted
correlation consistent Dunning basis sets~\cite{Dunning1989} augmented
with diffuse functions were tested extensively for the He atom,
especially to converge the lowest P state. No significant differences
were observed using the two basis sets and only the Dunning basis sets
are used in the following.  The basis sets used are: uncontracted
t-aug-cc-pV5Z for He, uncontracted d-aug-cc-pVDZ for Be, and
uncontracted d-aug-cc-pVTZ Dunning basis sets for H$_2$.

Calculations were performed for about 30 values of $\mu$ between $0$
to $10$ bohr$^{-1}$ (with about half the points between $0$ and $1$
where the energies vary the most). Cubic spline interpolation has been
used on this calculated data when plotting the total and excitation
energies as a function of $\mu$.  For later use, analytical
expressions were also fitted to the calculated total energies and
excitation energies. The forms used in the fitting were chosen to
satisfy the expansions at small and large $\mu$ values as presented in
Eqs.~\eqref{eq:Ei(mu=0)} and~\eqref{eq:E0infty}. The details of these
fits are given in the supplementary
material~\cite{RebTouTeaHelSav-JJJ-XX-sup}.

\section{Results and discussion}
\label{sec:results}

\subsection{Helium atom}

\begin{figure}
  \hspace*{-2.1cm}\begingroup
  \makeatletter
  \providecommand\color[2][]{    \GenericError{(gnuplot) \space\space\space\@spaces}{      Package color not loaded in conjunction with
      terminal option `colourtext'    }{See the gnuplot documentation for explanation.    }{Either use 'blacktext' in gnuplot or load the package
      color.sty in LaTeX.}    \renewcommand\color[2][]{}  }  \providecommand\includegraphics[2][]{    \GenericError{(gnuplot) \space\space\space\@spaces}{      Package graphicx or graphics not loaded    }{See the gnuplot documentation for explanation.    }{The gnuplot epslatex terminal needs graphicx.sty or graphics.sty.}    \renewcommand\includegraphics[2][]{}  }  \providecommand\rotatebox[2]{#2}  \@ifundefined{ifGPcolor}{    \newif\ifGPcolor
    \GPcolortrue
  }{}  \@ifundefined{ifGPblacktext}{    \newif\ifGPblacktext
    \GPblacktexttrue
  }{}    \let\gplgaddtomacro\g@addto@macro
    \gdef\gplbacktext{}  \gdef\gplfronttext{}  \makeatother
  \ifGPblacktext
        \def\colorrgb#1{}    \def\colorgray#1{}  \else
        \ifGPcolor
      \def\colorrgb#1{\color[rgb]{#1}}      \def\colorgray#1{\color[gray]{#1}}      \expandafter\def\csname LTw\endcsname{\color{white}}      \expandafter\def\csname LTb\endcsname{\color{black}}      \expandafter\def\csname LTa\endcsname{\color{black}}      \expandafter\def\csname LT0\endcsname{\color[rgb]{1,0,0}}      \expandafter\def\csname LT1\endcsname{\color[rgb]{0,1,0}}      \expandafter\def\csname LT2\endcsname{\color[rgb]{0,0,1}}      \expandafter\def\csname LT3\endcsname{\color[rgb]{1,0,1}}      \expandafter\def\csname LT4\endcsname{\color[rgb]{0,1,1}}      \expandafter\def\csname LT5\endcsname{\color[rgb]{1,1,0}}      \expandafter\def\csname LT6\endcsname{\color[rgb]{0,0,0}}      \expandafter\def\csname LT7\endcsname{\color[rgb]{1,0.3,0}}      \expandafter\def\csname LT8\endcsname{\color[rgb]{0.5,0.5,0.5}}    \else
            \def\colorrgb#1{\color{black}}      \def\colorgray#1{\color[gray]{#1}}      \expandafter\def\csname LTw\endcsname{\color{white}}      \expandafter\def\csname LTb\endcsname{\color{black}}      \expandafter\def\csname LTa\endcsname{\color{black}}      \expandafter\def\csname LT0\endcsname{\color{black}}      \expandafter\def\csname LT1\endcsname{\color{black}}      \expandafter\def\csname LT2\endcsname{\color{black}}      \expandafter\def\csname LT3\endcsname{\color{black}}      \expandafter\def\csname LT4\endcsname{\color{black}}      \expandafter\def\csname LT5\endcsname{\color{black}}      \expandafter\def\csname LT6\endcsname{\color{black}}      \expandafter\def\csname LT7\endcsname{\color{black}}      \expandafter\def\csname LT8\endcsname{\color{black}}    \fi
  \fi
  \setlength{\unitlength}{0.0500bp}  \begin{picture}(5760.00,4032.00)    \gplgaddtomacro\gplbacktext{      \csname LTb\endcsname      \put(1447,704){\makebox(0,0)[r]{\strut{}-3}}      \put(1447,1470){\makebox(0,0)[r]{\strut{}-2.5}}      \put(1447,2236){\makebox(0,0)[r]{\strut{}-2}}      \put(1447,3001){\makebox(0,0)[r]{\strut{}-1.5}}      \put(1447,3767){\makebox(0,0)[r]{\strut{}-1}}      \put(1579,484){\makebox(0,0){\strut{} 0}}      \put(1885,484){\makebox(0,0){\strut{} 1}}      \put(2192,484){\makebox(0,0){\strut{} 2}}      \put(2498,484){\makebox(0,0){\strut{} 3}}      \put(2804,484){\makebox(0,0){\strut{} 4}}      \put(3111,484){\makebox(0,0){\strut{} 5}}      \put(3417,484){\makebox(0,0){\strut{} 6}}      \put(3723,484){\makebox(0,0){\strut{} 7}}      \put(4029,484){\makebox(0,0){\strut{} 8}}      \put(4336,484){\makebox(0,0){\strut{} 9}}      \put(4642,484){\makebox(0,0){\strut{} 10}}      \put(3110,154){\makebox(0,0){\strut{}$\mu$ in bohr$^{-1}$}}      \put(966,2236){\rotatebox{90}{\makebox(0,0){\strut{}Total energies in hartree}}}    }    \gplgaddtomacro\gplfronttext{      \csname LTb\endcsname      \put(3586,3572){\makebox(0,0)[l]{\strut{}$\mathcal{E}_{\text{2S}} - \frac{2}{\sqrt{\pi}}\mu$}}      \csname LTb\endcsname      \put(3586,3308){\makebox(0,0)[l]{\strut{}$ 1 ^1 \text{S} $}}      \csname LTb\endcsname      \put(3586,3044){\makebox(0,0)[l]{\strut{}$2 ^3 \text{S} $}}      \csname LTb\endcsname      \put(3586,2780){\makebox(0,0)[l]{\strut{}$2 ^1 \text{S} $}}      \csname LTb\endcsname      \put(3586,2516){\makebox(0,0)[l]{\strut{}$1 ^3 \text{P} $}}      \csname LTb\endcsname      \put(3586,2252){\makebox(0,0)[l]{\strut{}$1 ^1 \text{P} $}}    }    \gplbacktext
    \put(0,0){\includegraphics{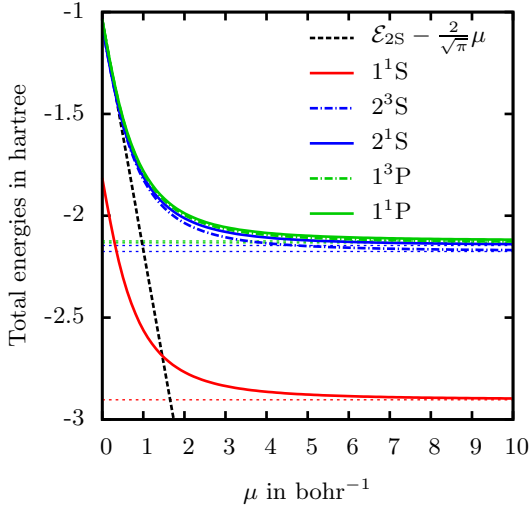}}    \gplfronttext
  \end{picture}\endgroup
  \caption{Ground- and excited-state total energies ${\cal E}_k^{\mu}$ (in hartree) of the He atom as a function of $\mu$ (in bohr$^{-1}$). The total energies of the physical system $E_k = {\cal E}_k^{\mu\to\infty}$ are plotted as horizontal dotted lines. The slope at $\mu=0$ is shown by the black dashed line for the first excited state.
 \label{fig:he_tacpv5z_state_svd07_0}
  }
\end{figure}

\begin{figure}
  \hspace*{-2.1cm}
\begingroup
  \makeatletter
  \providecommand\color[2][]{    \GenericError{(gnuplot) \space\space\space\@spaces}{      Package color not loaded in conjunction with
      terminal option `colourtext'    }{See the gnuplot documentation for explanation.    }{Either use 'blacktext' in gnuplot or load the package
      color.sty in LaTeX.}    \renewcommand\color[2][]{}  }  \providecommand\includegraphics[2][]{    \GenericError{(gnuplot) \space\space\space\@spaces}{      Package graphicx or graphics not loaded    }{See the gnuplot documentation for explanation.    }{The gnuplot epslatex terminal needs graphicx.sty or graphics.sty.}    \renewcommand\includegraphics[2][]{}  }  \providecommand\rotatebox[2]{#2}  \@ifundefined{ifGPcolor}{    \newif\ifGPcolor
    \GPcolortrue
  }{}  \@ifundefined{ifGPblacktext}{    \newif\ifGPblacktext
    \GPblacktexttrue
  }{}    \let\gplgaddtomacro\g@addto@macro
    \gdef\gplbacktext{}  \gdef\gplfronttext{}  \makeatother
  \ifGPblacktext
        \def\colorrgb#1{}    \def\colorgray#1{}  \else
        \ifGPcolor
      \def\colorrgb#1{\color[rgb]{#1}}      \def\colorgray#1{\color[gray]{#1}}      \expandafter\def\csname LTw\endcsname{\color{white}}      \expandafter\def\csname LTb\endcsname{\color{black}}      \expandafter\def\csname LTa\endcsname{\color{black}}      \expandafter\def\csname LT0\endcsname{\color[rgb]{1,0,0}}      \expandafter\def\csname LT1\endcsname{\color[rgb]{0,1,0}}      \expandafter\def\csname LT2\endcsname{\color[rgb]{0,0,1}}      \expandafter\def\csname LT3\endcsname{\color[rgb]{1,0,1}}      \expandafter\def\csname LT4\endcsname{\color[rgb]{0,1,1}}      \expandafter\def\csname LT5\endcsname{\color[rgb]{1,1,0}}      \expandafter\def\csname LT6\endcsname{\color[rgb]{0,0,0}}      \expandafter\def\csname LT7\endcsname{\color[rgb]{1,0.3,0}}      \expandafter\def\csname LT8\endcsname{\color[rgb]{0.5,0.5,0.5}}    \else
            \def\colorrgb#1{\color{black}}      \def\colorgray#1{\color[gray]{#1}}      \expandafter\def\csname LTw\endcsname{\color{white}}      \expandafter\def\csname LTb\endcsname{\color{black}}      \expandafter\def\csname LTa\endcsname{\color{black}}      \expandafter\def\csname LT0\endcsname{\color{black}}      \expandafter\def\csname LT1\endcsname{\color{black}}      \expandafter\def\csname LT2\endcsname{\color{black}}      \expandafter\def\csname LT3\endcsname{\color{black}}      \expandafter\def\csname LT4\endcsname{\color{black}}      \expandafter\def\csname LT5\endcsname{\color{black}}      \expandafter\def\csname LT6\endcsname{\color{black}}      \expandafter\def\csname LT7\endcsname{\color{black}}      \expandafter\def\csname LT8\endcsname{\color{black}}    \fi
  \fi
  \setlength{\unitlength}{0.0500bp}  \begin{picture}(5760.00,4032.00)    \gplgaddtomacro\gplbacktext{      \csname LTb\endcsname      \put(1513,704){\makebox(0,0)[r]{\strut{} 0.72}}      \put(1513,1142){\makebox(0,0)[r]{\strut{} 0.73}}      \put(1513,1579){\makebox(0,0)[r]{\strut{} 0.74}}      \put(1513,2017){\makebox(0,0)[r]{\strut{} 0.75}}      \put(1513,2454){\makebox(0,0)[r]{\strut{} 0.76}}      \put(1513,2892){\makebox(0,0)[r]{\strut{} 0.77}}      \put(1513,3329){\makebox(0,0)[r]{\strut{} 0.78}}      \put(1513,3767){\makebox(0,0)[r]{\strut{} 0.79}}      \put(1645,484){\makebox(0,0){\strut{} 0}}      \put(2258,484){\makebox(0,0){\strut{} 1}}      \put(2870,484){\makebox(0,0){\strut{} 2}}      \put(3483,484){\makebox(0,0){\strut{} 3}}      \put(4095,484){\makebox(0,0){\strut{} 4}}      \put(4708,484){\makebox(0,0){\strut{} 5}}      \put(3176,154){\makebox(0,0){\strut{}$\mu$ in bohr$^{-1}$}}      \put(971,2236){\rotatebox{90}{\makebox(0,0){\strut{}Excitation energies in hartree}}}    }    \gplgaddtomacro\gplfronttext{      \csname LTb\endcsname      \put(3675,2104){\makebox(0,0)[l]{\strut{}$1^1\text{S} \to 2^3\text{S} $}}      \csname LTb\endcsname      \put(3675,1840){\makebox(0,0)[l]{\strut{}$1^1\text{S} \to 2^1\text{S} $}}      \csname LTb\endcsname      \put(3675,1576){\makebox(0,0)[l]{\strut{}$1^1\text{S} \to 1^3\text{P} $}}      \csname LTb\endcsname      \put(3675,1312){\makebox(0,0)[l]{\strut{}$1^1\text{S} \to 1^1\text{P} $}}    }    \gplbacktext
    \put(0,0){\includegraphics{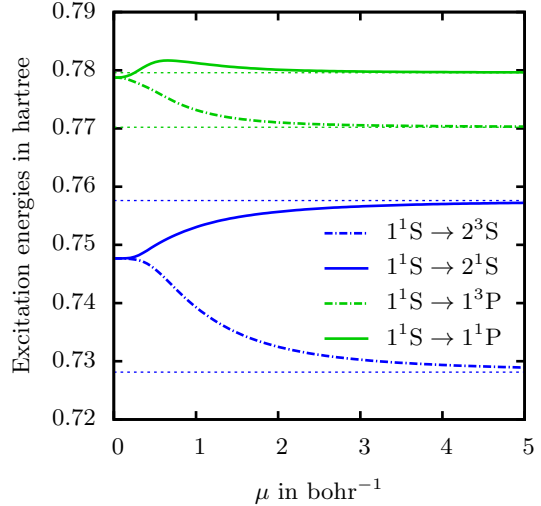}}    \gplfronttext
  \end{picture}\endgroup
  \caption{Excitation energies $\Delta {\cal E}_k^{\mu} = {\cal E}_k^{\mu} - {\cal E}_0^{\mu}$ (in hartree) of the He atom as a function of $\mu$ (in bohr$^{-1}$). The excitation energies of the physical system $\Delta E_k = \Delta {\cal E}_k^{\mu\to\infty}$ are plotted as horizontal dotted lines.
    \label{fig:he_exc_tacpv5z_svd07_0}
  }
\end{figure}

The total energies of the ground state $1^1$S and of the first
Rydberg-like singlet and triplet S and P excited states of the He atom
are plotted as a function of the range-separation parameter $\mu$ in
Figure~\ref{fig:he_tacpv5z_state_svd07_0}. At $\mu=0$, the KS
non-interacting total energies are obtained.  Being sums of orbital
energies with a resulting double counting of electron repulsion, these
quantities are well above the total energies of the physical system
(higher by about 1 hartree).  When the long-range electron--electron
interaction is added by increasing $\mu$ from $\mu=0$, the total
energies decrease linearly with $\mu$ with a slope of $-2/\sqrt{\pi}$,
in accordance with the linear term in the expansion of
Eq.\;(\ref{eq:Ei(mu=0)}) for $N=2$.  For larger $\mu$ values, the
total energy curves flatten and approach the energies of the physical
system asymptotically as $1/\mu^2$ as $\mu \to \infty$, in accordance
with Eq.\;(\ref{eq:E0infty}). The total energies along the adiabatic
connection are poor approximations to the total energies of the
physical system unless the range-separation parameter $\mu$ is
large. Specifically, $\mu \gtrsim 6$ is required to be within
10\;mhartree of the exact total energies.

The lowest singlet and triplet excitation energies are plotted in
Figure\;\ref{fig:he_exc_tacpv5z_svd07_0}.  The KS singlet and triplet
excitation energies are degenerate and, as already observed for a few
atomic systems in
Refs.\;\onlinecite{UmrSavGon-INC-98,Filippi1997,Savin1998}, are
bracketed by the singlet and triplet excitation energies of the
physical system.  As $\mu$ increases from $\mu=0$, the excitation
energies vary as $\mu^3$ since the linear term in
Eq.\;(\ref{eq:Ei(mu=0)}) cancels out for energy differences. The
singlet--triplet degeneracy is lifted and the excitation energies
converge to the exact singlet and triplet excitation energies when
$\mu\to\infty$.  Whereas a monotonic variation of the excitation
energy with $\mu$ can be observed for the singlet and triplet 1S$ \to$
2S excitations and for the triplet 1$^1$S $\to$ 1$^3$P excitation, a
non-monotonic variation is observed for the singlet 1$^1$S $\to$
1$^1$P excitation.  This behaviour could be an artefact of the
basis-set expansions (either orbital or potential), noting that a
similar behaviour was observed for other excitations in a smaller
basis set and was removed by enlarging the basis set (the basis set
dependence of the singlet 1$^1$S $\to$ 1$^1$P excitation energy is
given in the supplementary
material~\cite{RebTouTeaHelSav-JJJ-XX-sup}).  In line with previous
observations in Refs.~\cite{UmrSavGon-INC-98,Savin1998} for the KS
system, the excitation energies for Rydberg-type states along the
adiabatic connection are rather good approximations to the excitation
energies of the physical system (the maximal error is about 0.02
hartree at $\mu=0$ for the triplet 1$^1$S $\to$ 2$^3$S excitation),
becoming better and better for high-lying states as they must
eventually converge to the exact ionization energy.

\begin{figure}
  \hspace*{-2.1cm}
\begingroup
  \makeatletter
  \providecommand\color[2][]{    \GenericError{(gnuplot) \space\space\space\@spaces}{      Package color not loaded in conjunction with
      terminal option `colourtext'    }{See the gnuplot documentation for explanation.    }{Either use 'blacktext' in gnuplot or load the package
      color.sty in LaTeX.}    \renewcommand\color[2][]{}  }  \providecommand\includegraphics[2][]{    \GenericError{(gnuplot) \space\space\space\@spaces}{      Package graphicx or graphics not loaded    }{See the gnuplot documentation for explanation.    }{The gnuplot epslatex terminal needs graphicx.sty or graphics.sty.}    \renewcommand\includegraphics[2][]{}  }  \providecommand\rotatebox[2]{#2}  \@ifundefined{ifGPcolor}{    \newif\ifGPcolor
    \GPcolortrue
  }{}  \@ifundefined{ifGPblacktext}{    \newif\ifGPblacktext
    \GPblacktexttrue
  }{}    \let\gplgaddtomacro\g@addto@macro
    \gdef\gplbacktext{}  \gdef\gplfronttext{}  \makeatother
  \ifGPblacktext
        \def\colorrgb#1{}    \def\colorgray#1{}  \else
        \ifGPcolor
      \def\colorrgb#1{\color[rgb]{#1}}      \def\colorgray#1{\color[gray]{#1}}      \expandafter\def\csname LTw\endcsname{\color{white}}      \expandafter\def\csname LTb\endcsname{\color{black}}      \expandafter\def\csname LTa\endcsname{\color{black}}      \expandafter\def\csname LT0\endcsname{\color[rgb]{1,0,0}}      \expandafter\def\csname LT1\endcsname{\color[rgb]{0,1,0}}      \expandafter\def\csname LT2\endcsname{\color[rgb]{0,0,1}}      \expandafter\def\csname LT3\endcsname{\color[rgb]{1,0,1}}      \expandafter\def\csname LT4\endcsname{\color[rgb]{0,1,1}}      \expandafter\def\csname LT5\endcsname{\color[rgb]{1,1,0}}      \expandafter\def\csname LT6\endcsname{\color[rgb]{0,0,0}}      \expandafter\def\csname LT7\endcsname{\color[rgb]{1,0.3,0}}      \expandafter\def\csname LT8\endcsname{\color[rgb]{0.5,0.5,0.5}}    \else
            \def\colorrgb#1{\color{black}}      \def\colorgray#1{\color[gray]{#1}}      \expandafter\def\csname LTw\endcsname{\color{white}}      \expandafter\def\csname LTb\endcsname{\color{black}}      \expandafter\def\csname LTa\endcsname{\color{black}}      \expandafter\def\csname LT0\endcsname{\color{black}}      \expandafter\def\csname LT1\endcsname{\color{black}}      \expandafter\def\csname LT2\endcsname{\color{black}}      \expandafter\def\csname LT3\endcsname{\color{black}}      \expandafter\def\csname LT4\endcsname{\color{black}}      \expandafter\def\csname LT5\endcsname{\color{black}}      \expandafter\def\csname LT6\endcsname{\color{black}}      \expandafter\def\csname LT7\endcsname{\color{black}}      \expandafter\def\csname LT8\endcsname{\color{black}}    \fi
  \fi
  \setlength{\unitlength}{0.0500bp}  \begin{picture}(5760.00,4032.00)    \gplgaddtomacro\gplbacktext{      \csname LTb\endcsname      \put(1579,704){\makebox(0,0)[r]{\strut{} 0}}      \put(1579,1725){\makebox(0,0)[r]{\strut{} 0.005}}      \put(1579,2746){\makebox(0,0)[r]{\strut{} 0.01}}      \put(1579,3767){\makebox(0,0)[r]{\strut{} 0.015}}      \put(1711,484){\makebox(0,0){\strut{} 0}}      \put(2324,484){\makebox(0,0){\strut{} 0.2}}      \put(2936,484){\makebox(0,0){\strut{} 0.4}}      \put(3549,484){\makebox(0,0){\strut{} 0.6}}      \put(4161,484){\makebox(0,0){\strut{} 0.8}}      \put(4774,484){\makebox(0,0){\strut{} 1}}      \put(3242,154){\makebox(0,0){\strut{}$\mu$ in  bohr$^{-1}$}}      \put(1007,2133){\rotatebox{90}{\makebox(0,0){\strut{}Singlet-triplet energy splitting in hartree}}}    }    \gplgaddtomacro\gplfronttext{      \csname LTb\endcsname      \put(3982,3572){\makebox(0,0)[l]{\strut{}$2 \text{S} $}}      \csname LTb\endcsname      \put(3982,3308){\makebox(0,0)[l]{\strut{}$1 \text{P} $}}    }    \gplbacktext
    \put(0,0){\includegraphics{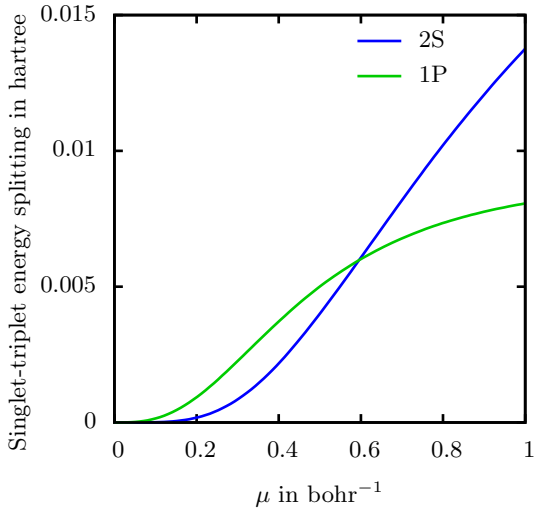}}    \gplfronttext
  \end{picture}\endgroup
  \caption{Singlet--triplet energy splittings (in hartree) for the He atom as a function of $\mu$ (in bohr$^{-1}$).
    \label{fig:he_singtrip_tacpv5z_dalton_svd07_0}
  }
\end{figure}

The singlet--triplet energy splittings for the 2S and 1P states are
plotted in Figure\;\ref{fig:he_singtrip_tacpv5z_dalton_svd07_0}.  The
expansion at small $\mu$ of Eq.~\eqref{eq:spin splitting} predicts the
singlet--triplet splitting to increase as $\mu^3$ for the 1P state
since it corresponds to the 1s $\to$ 2p excitation in the KS system,
so that $\Delta \ell =1$.  By contrast, the singlet--triplet splitting
should increase at most as $\mu^5$ for the 2S state since it
corresponds to the 1s $\to$ 2s excitation in the KS system, so that
$\Delta \ell =0$.  This difference is clearly visible in
Figure~\ref{fig:he_singtrip_tacpv5z_dalton_svd07_0}, where the 2S
curve for the singlet--triplet splitting initially increases more
slowly than the 1P curve.

\subsection{Beryllium atom}

\begin{figure}
  \hspace*{-2.1cm}
\begingroup
  \makeatletter
  \providecommand\color[2][]{    \GenericError{(gnuplot) \space\space\space\@spaces}{      Package color not loaded in conjunction with
      terminal option `colourtext'    }{See the gnuplot documentation for explanation.    }{Either use 'blacktext' in gnuplot or load the package
      color.sty in LaTeX.}    \renewcommand\color[2][]{}  }  \providecommand\includegraphics[2][]{    \GenericError{(gnuplot) \space\space\space\@spaces}{      Package graphicx or graphics not loaded    }{See the gnuplot documentation for explanation.    }{The gnuplot epslatex terminal needs graphicx.sty or graphics.sty.}    \renewcommand\includegraphics[2][]{}  }  \providecommand\rotatebox[2]{#2}  \@ifundefined{ifGPcolor}{    \newif\ifGPcolor
    \GPcolortrue
  }{}  \@ifundefined{ifGPblacktext}{    \newif\ifGPblacktext
    \GPblacktexttrue
  }{}    \let\gplgaddtomacro\g@addto@macro
    \gdef\gplbacktext{}  \gdef\gplfronttext{}  \makeatother
  \ifGPblacktext
        \def\colorrgb#1{}    \def\colorgray#1{}  \else
        \ifGPcolor
      \def\colorrgb#1{\color[rgb]{#1}}      \def\colorgray#1{\color[gray]{#1}}      \expandafter\def\csname LTw\endcsname{\color{white}}      \expandafter\def\csname LTb\endcsname{\color{black}}      \expandafter\def\csname LTa\endcsname{\color{black}}      \expandafter\def\csname LT0\endcsname{\color[rgb]{1,0,0}}      \expandafter\def\csname LT1\endcsname{\color[rgb]{0,1,0}}      \expandafter\def\csname LT2\endcsname{\color[rgb]{0,0,1}}      \expandafter\def\csname LT3\endcsname{\color[rgb]{1,0,1}}      \expandafter\def\csname LT4\endcsname{\color[rgb]{0,1,1}}      \expandafter\def\csname LT5\endcsname{\color[rgb]{1,1,0}}      \expandafter\def\csname LT6\endcsname{\color[rgb]{0,0,0}}      \expandafter\def\csname LT7\endcsname{\color[rgb]{1,0.3,0}}      \expandafter\def\csname LT8\endcsname{\color[rgb]{0.5,0.5,0.5}}    \else
            \def\colorrgb#1{\color{black}}      \def\colorgray#1{\color[gray]{#1}}      \expandafter\def\csname LTw\endcsname{\color{white}}      \expandafter\def\csname LTb\endcsname{\color{black}}      \expandafter\def\csname LTa\endcsname{\color{black}}      \expandafter\def\csname LT0\endcsname{\color{black}}      \expandafter\def\csname LT1\endcsname{\color{black}}      \expandafter\def\csname LT2\endcsname{\color{black}}      \expandafter\def\csname LT3\endcsname{\color{black}}      \expandafter\def\csname LT4\endcsname{\color{black}}      \expandafter\def\csname LT5\endcsname{\color{black}}      \expandafter\def\csname LT6\endcsname{\color{black}}      \expandafter\def\csname LT7\endcsname{\color{black}}      \expandafter\def\csname LT8\endcsname{\color{black}}    \fi
  \fi
  \setlength{\unitlength}{0.0500bp}  \begin{picture}(5760.00,4032.00)    \gplgaddtomacro\gplbacktext{      \csname LTb\endcsname      \put(1381,704){\makebox(0,0)[r]{\strut{}-15}}      \put(1381,1142){\makebox(0,0)[r]{\strut{}-14}}      \put(1381,1579){\makebox(0,0)[r]{\strut{}-13}}      \put(1381,2017){\makebox(0,0)[r]{\strut{}-12}}      \put(1381,2454){\makebox(0,0)[r]{\strut{}-11}}      \put(1381,2892){\makebox(0,0)[r]{\strut{}-10}}      \put(1381,3329){\makebox(0,0)[r]{\strut{}-9}}      \put(1381,3767){\makebox(0,0)[r]{\strut{}-8}}      \put(1513,484){\makebox(0,0){\strut{} 0}}      \put(2126,484){\makebox(0,0){\strut{} 2}}      \put(2738,484){\makebox(0,0){\strut{} 4}}      \put(3351,484){\makebox(0,0){\strut{} 6}}      \put(3963,484){\makebox(0,0){\strut{} 8}}      \put(4576,484){\makebox(0,0){\strut{} 10}}      \put(3044,154){\makebox(0,0){\strut{}$\mu$ in bohr$^{-1}$}}      \put(900,2236){\rotatebox{90}{\makebox(0,0){\strut{}Total energies in hartree}}}    }    \gplgaddtomacro\gplfronttext{      \csname LTb\endcsname      \put(3520,3572){\makebox(0,0)[l]{\strut{}$\mathcal{E}_{\text{1S}} - \frac{12}{\sqrt{\pi}}\mu$}}      \csname LTb\endcsname      \put(3520,3308){\makebox(0,0)[l]{\strut{}$ 1 ^1 \text{S} $}}      \csname LTb\endcsname      \put(3520,3044){\makebox(0,0)[l]{\strut{}$ 1 ^3 \text{P} $}}      \csname LTb\endcsname      \put(3520,2780){\makebox(0,0)[l]{\strut{}$ 1 ^1 \text{P} $}}    }    \gplbacktext
    \put(0,0){\includegraphics{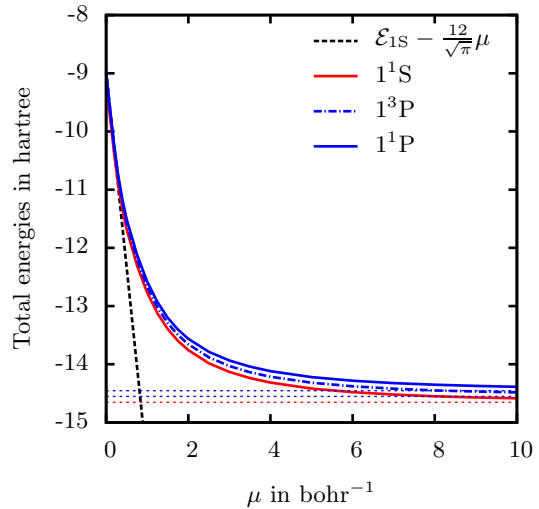}}    \gplfronttext
  \end{picture}\endgroup
  \caption{Ground- and excited-state total energies ${\cal E}_k^{\mu}$ (in hartree) of the Be atom as a function of $\mu$ (in bohr$^{-1}$). The total energies of the physical system $E_k = {\cal E}_k^{\mu\to\infty}$ are plotted as horizontal dotted lines. The slope at $\mu=0$ is shown in dashed line.
    \label{fig:be_dacpvdz_state_0}
  }
\end{figure}

\begin{figure}
  \hspace*{-2.1cm}
\begingroup
  \makeatletter
  \providecommand\color[2][]{    \GenericError{(gnuplot) \space\space\space\@spaces}{      Package color not loaded in conjunction with
      terminal option `colourtext'    }{See the gnuplot documentation for explanation.    }{Either use 'blacktext' in gnuplot or load the package
      color.sty in LaTeX.}    \renewcommand\color[2][]{}  }  \providecommand\includegraphics[2][]{    \GenericError{(gnuplot) \space\space\space\@spaces}{      Package graphicx or graphics not loaded    }{See the gnuplot documentation for explanation.    }{The gnuplot epslatex terminal needs graphicx.sty or graphics.sty.}    \renewcommand\includegraphics[2][]{}  }  \providecommand\rotatebox[2]{#2}  \@ifundefined{ifGPcolor}{    \newif\ifGPcolor
    \GPcolortrue
  }{}  \@ifundefined{ifGPblacktext}{    \newif\ifGPblacktext
    \GPblacktexttrue
  }{}    \let\gplgaddtomacro\g@addto@macro
    \gdef\gplbacktext{}  \gdef\gplfronttext{}  \makeatother
  \ifGPblacktext
        \def\colorrgb#1{}    \def\colorgray#1{}  \else
        \ifGPcolor
      \def\colorrgb#1{\color[rgb]{#1}}      \def\colorgray#1{\color[gray]{#1}}      \expandafter\def\csname LTw\endcsname{\color{white}}      \expandafter\def\csname LTb\endcsname{\color{black}}      \expandafter\def\csname LTa\endcsname{\color{black}}      \expandafter\def\csname LT0\endcsname{\color[rgb]{1,0,0}}      \expandafter\def\csname LT1\endcsname{\color[rgb]{0,1,0}}      \expandafter\def\csname LT2\endcsname{\color[rgb]{0,0,1}}      \expandafter\def\csname LT3\endcsname{\color[rgb]{1,0,1}}      \expandafter\def\csname LT4\endcsname{\color[rgb]{0,1,1}}      \expandafter\def\csname LT5\endcsname{\color[rgb]{1,1,0}}      \expandafter\def\csname LT6\endcsname{\color[rgb]{0,0,0}}      \expandafter\def\csname LT7\endcsname{\color[rgb]{1,0.3,0}}      \expandafter\def\csname LT8\endcsname{\color[rgb]{0.5,0.5,0.5}}    \else
            \def\colorrgb#1{\color{black}}      \def\colorgray#1{\color[gray]{#1}}      \expandafter\def\csname LTw\endcsname{\color{white}}      \expandafter\def\csname LTb\endcsname{\color{black}}      \expandafter\def\csname LTa\endcsname{\color{black}}      \expandafter\def\csname LT0\endcsname{\color{black}}      \expandafter\def\csname LT1\endcsname{\color{black}}      \expandafter\def\csname LT2\endcsname{\color{black}}      \expandafter\def\csname LT3\endcsname{\color{black}}      \expandafter\def\csname LT4\endcsname{\color{black}}      \expandafter\def\csname LT5\endcsname{\color{black}}      \expandafter\def\csname LT6\endcsname{\color{black}}      \expandafter\def\csname LT7\endcsname{\color{black}}      \expandafter\def\csname LT8\endcsname{\color{black}}    \fi
  \fi
  \setlength{\unitlength}{0.0500bp}  \begin{picture}(5760.00,4032.00)    \gplgaddtomacro\gplbacktext{      \csname LTb\endcsname      \put(1513,704){\makebox(0,0)[r]{\strut{} 0.08}}      \put(1513,1142){\makebox(0,0)[r]{\strut{} 0.1}}      \put(1513,1579){\makebox(0,0)[r]{\strut{} 0.12}}      \put(1513,2017){\makebox(0,0)[r]{\strut{} 0.14}}      \put(1513,2454){\makebox(0,0)[r]{\strut{} 0.16}}      \put(1513,2892){\makebox(0,0)[r]{\strut{} 0.18}}      \put(1513,3329){\makebox(0,0)[r]{\strut{} 0.2}}      \put(1513,3767){\makebox(0,0)[r]{\strut{} 0.22}}      \put(1645,484){\makebox(0,0){\strut{} 0}}      \put(2258,484){\makebox(0,0){\strut{} 1}}      \put(2870,484){\makebox(0,0){\strut{} 2}}      \put(3483,484){\makebox(0,0){\strut{} 3}}      \put(4095,484){\makebox(0,0){\strut{} 4}}      \put(4708,484){\makebox(0,0){\strut{} 5}}      \put(3176,154){\makebox(0,0){\strut{}$\mu$ in bohr$^{-1}$}}      \put(971,2236){\rotatebox{90}{\makebox(0,0){\strut{}Excitation energies in hartree}}}    }    \gplgaddtomacro\gplfronttext{      \csname LTb\endcsname      \put(3581,2541){\makebox(0,0)[l]{\strut{}$ 1 ^1 \text{S} \to 1 ^3 \text{P} $}}      \csname LTb\endcsname      \put(3581,2277){\makebox(0,0)[l]{\strut{}$ 1 ^1 \text{S} \to 1 ^1 \text{P} $}}    }    \gplbacktext
    \put(0,0){\includegraphics{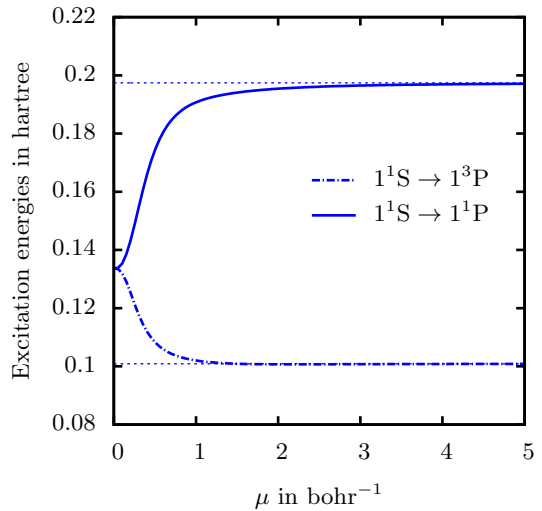}}    \gplfronttext
  \end{picture}\endgroup
  \caption{Excitation energies $\Delta {\cal E}_k^{\mu} = {\cal E}_k^{\mu} - {\cal E}_0^{\mu}$ (in hartree) of the Be atom as a function of $\mu$ (in bohr$^{-1}$). The excitation energies of the physical system $\Delta E_k = \Delta {\cal E}_k^{\mu\to\infty}$ are plotted as horizontal dotted lines.
    \label{fig:be_dacpvdz_exc_0}
  }
\end{figure}

The total energies of the ground state 1$^1$S and of the valence
singlet and triplet 1P excited states of the Be atom are plotted in
Figure\;\ref{fig:be_dacpvdz_state_0}. The KS total energies are
approximately $6$ hartree above the physical energies.  At small
$\mu$, an initial slope of $-12/\sqrt{\pi}$ is observed for all
states, in accordance with Eq.\;\eqref{eq:Ei(mu=0)} with $N=4$.
However, convergence to the physical energies with increasing $\mu$ is
much slower than for the He atom, owing to the short inter-electronic
distances in the Be 1s core region, which are consequently probed at
larger $\mu$ values.

The singlet and triplet excitation energies are plotted in
Figure\;\ref{fig:be_dacpvdz_exc_0}.  As for He, the KS excitation
energies are bracketed by the singlet and triplet excitation energies
of the physical system.  Not surprisingly, the KS excitation energies
are poorer approximations to the exact excitation energies for these
valence excitations in Be than for the Rydberg excitations in He.  As
$\mu$ increases, the KS excitation energies rapidly converge to the
physical excitation energies.  Clearly, the slow convergence of the
core energies does not affect the convergence of the valence
excitation energies.

Close to the KS system, at $\mu=0$, the excitation energies are quite
sensitive to the introduction of a small portion of electron--electron
interaction in the Hamiltonian, which may be interpreted as a sign of
static correlation.  For $\mu \approx 0.4-0.5$, a typical $\mu$ value
in range-separated DFT
calculations~\cite{GerAng-CPL-05,FroTouJen-JCP-07}, the calculated
excitation energies are significantly better approximations to the
exact excitation energies than are the KS excitation energies.  This
observation justifies range-separated multi-determinantal
linear-response DFT calculations, which take these excitation energies
as a starting point.

\subsection{Hydrogen molecule}

\begin{figure}
  \hspace*{-2.1cm}
\begingroup
  \makeatletter
  \providecommand\color[2][]{    \GenericError{(gnuplot) \space\space\space\@spaces}{      Package color not loaded in conjunction with
      terminal option `colourtext'    }{See the gnuplot documentation for explanation.    }{Either use 'blacktext' in gnuplot or load the package
      color.sty in LaTeX.}    \renewcommand\color[2][]{}  }  \providecommand\includegraphics[2][]{    \GenericError{(gnuplot) \space\space\space\@spaces}{      Package graphicx or graphics not loaded    }{See the gnuplot documentation for explanation.    }{The gnuplot epslatex terminal needs graphicx.sty or graphics.sty.}    \renewcommand\includegraphics[2][]{}  }  \providecommand\rotatebox[2]{#2}  \@ifundefined{ifGPcolor}{    \newif\ifGPcolor
    \GPcolortrue
  }{}  \@ifundefined{ifGPblacktext}{    \newif\ifGPblacktext
    \GPblacktexttrue
  }{}    \let\gplgaddtomacro\g@addto@macro
    \gdef\gplbacktext{}  \gdef\gplfronttext{}  \makeatother
  \ifGPblacktext
        \def\colorrgb#1{}    \def\colorgray#1{}  \else
        \ifGPcolor
      \def\colorrgb#1{\color[rgb]{#1}}      \def\colorgray#1{\color[gray]{#1}}      \expandafter\def\csname LTw\endcsname{\color{white}}      \expandafter\def\csname LTb\endcsname{\color{black}}      \expandafter\def\csname LTa\endcsname{\color{black}}      \expandafter\def\csname LT0\endcsname{\color[rgb]{1,0,0}}      \expandafter\def\csname LT1\endcsname{\color[rgb]{0,1,0}}      \expandafter\def\csname LT2\endcsname{\color[rgb]{0,0,1}}      \expandafter\def\csname LT3\endcsname{\color[rgb]{1,0,1}}      \expandafter\def\csname LT4\endcsname{\color[rgb]{0,1,1}}      \expandafter\def\csname LT5\endcsname{\color[rgb]{1,1,0}}      \expandafter\def\csname LT6\endcsname{\color[rgb]{0,0,0}}      \expandafter\def\csname LT7\endcsname{\color[rgb]{1,0.3,0}}      \expandafter\def\csname LT8\endcsname{\color[rgb]{0.5,0.5,0.5}}    \else
            \def\colorrgb#1{\color{black}}      \def\colorgray#1{\color[gray]{#1}}      \expandafter\def\csname LTw\endcsname{\color{white}}      \expandafter\def\csname LTb\endcsname{\color{black}}      \expandafter\def\csname LTa\endcsname{\color{black}}      \expandafter\def\csname LT0\endcsname{\color{black}}      \expandafter\def\csname LT1\endcsname{\color{black}}      \expandafter\def\csname LT2\endcsname{\color{black}}      \expandafter\def\csname LT3\endcsname{\color{black}}      \expandafter\def\csname LT4\endcsname{\color{black}}      \expandafter\def\csname LT5\endcsname{\color{black}}      \expandafter\def\csname LT6\endcsname{\color{black}}      \expandafter\def\csname LT7\endcsname{\color{black}}      \expandafter\def\csname LT8\endcsname{\color{black}}    \fi
  \fi
  \setlength{\unitlength}{0.0500bp}  \begin{picture}(5760.00,4032.00)    \gplgaddtomacro\gplbacktext{      \csname LTb\endcsname      \put(1513,704){\makebox(0,0)[r]{\strut{} 0.38}}      \put(1513,1215){\makebox(0,0)[r]{\strut{} 0.4}}      \put(1513,1725){\makebox(0,0)[r]{\strut{} 0.42}}      \put(1513,2236){\makebox(0,0)[r]{\strut{} 0.44}}      \put(1513,2746){\makebox(0,0)[r]{\strut{} 0.46}}      \put(1513,3257){\makebox(0,0)[r]{\strut{} 0.48}}      \put(1513,3767){\makebox(0,0)[r]{\strut{} 0.5}}      \put(1645,484){\makebox(0,0){\strut{} 0}}      \put(2258,484){\makebox(0,0){\strut{} 1}}      \put(2870,484){\makebox(0,0){\strut{} 2}}      \put(3483,484){\makebox(0,0){\strut{} 3}}      \put(4095,484){\makebox(0,0){\strut{} 4}}      \put(4708,484){\makebox(0,0){\strut{} 5}}      \put(3176,154){\makebox(0,0){\strut{}$\mu$ in  bohr$^{-1}$}}      \put(971,2236){\rotatebox{90}{\makebox(0,0){\strut{}Excitation energies in hartree}}}    }    \gplgaddtomacro\gplfronttext{      \csname LTb\endcsname      \put(3388,2486){\makebox(0,0)[l]{\strut{}$1 ^1\Sigma_\text g^+ \to 1 ^3\Sigma_\text u^+ $}}      \csname LTb\endcsname      \put(3388,2222){\makebox(0,0)[l]{\strut{}$1 ^1\Sigma_\text g^+ \to 1 ^1\Sigma_\text u^+ $}}      \csname LTb\endcsname      \put(3388,1958){\makebox(0,0)[l]{\strut{}$1 ^1\Sigma_\text g^+ \to 2 ^3\Sigma_\text g^+ $}}      \csname LTb\endcsname      \put(3388,1694){\makebox(0,0)[l]{\strut{}$1 ^1\Sigma_\text g^+ \to 2 ^1\Sigma_\text g^+ $}}      \csname LTb\endcsname      \put(3388,1430){\makebox(0,0)[l]{\strut{}$1 ^1\Sigma_\text g^+ \to 1 ^3\Pi_\text u $}}      \csname LTb\endcsname      \put(3388,1166){\makebox(0,0)[l]{\strut{}$1 ^1\Sigma_\text g^+ \to 1 ^1\Pi_\text u  $}}    }    \gplbacktext
    \put(0,0){\includegraphics{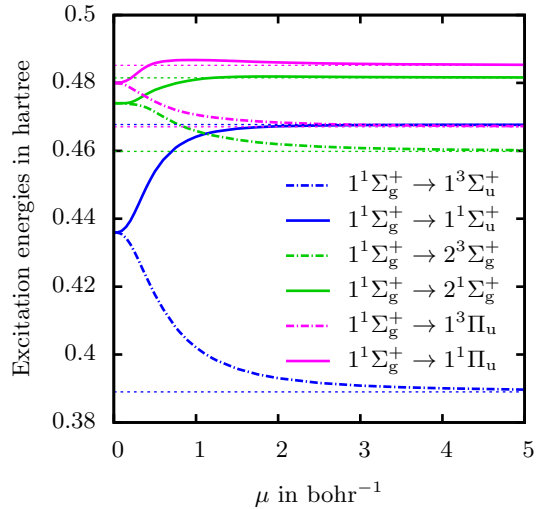}}    \gplfronttext
  \end{picture}\endgroup
  \caption{
Excitation energies $\Delta {\cal E}_k^{\mu} = {\cal E}_k^{\mu} - {\cal E}_0^{\mu}$ (in hartree) of the H$_2$ molecule at the equilibrium internuclear distance as a function of $\mu$ (in bohr$^{-1}$). The excitation energies of the physical system $\Delta E_k = \Delta {\cal E}_k^{\mu\to\infty}$ are plotted as horizontal dotted lines.
    \label{fig:h2_eq_dacpvtz_exc_0}
  }
\end{figure}

The first few excitation energies of H$_2$ at the equilibrium bond distance are plotted against $\mu$ in Figure\;\ref{fig:h2_eq_dacpvtz_exc_0}. 
As for the atoms, the valence excitations energies vary much more along the adiabatic connection than do the Rydberg-like excitation energies. 
Note also that the energetic ordering of the states changes along the adiabatic connection. 
With our choice of basis set, we also observe that the higher singlet excitation energies do not depend monotonically 
on $\mu$, approaching the physical limits from above, as observed for He. Again, the excitation energies around $\mu \approx 0.4-0.5$ represent better approximations to the exact excitation energies than the KS excitation energies.

\begin{figure}
  \hspace*{-2.1cm}
\begingroup
  \makeatletter
  \providecommand\color[2][]{    \GenericError{(gnuplot) \space\space\space\@spaces}{      Package color not loaded in conjunction with
      terminal option `colourtext'    }{See the gnuplot documentation for explanation.    }{Either use 'blacktext' in gnuplot or load the package
      color.sty in LaTeX.}    \renewcommand\color[2][]{}  }  \providecommand\includegraphics[2][]{    \GenericError{(gnuplot) \space\space\space\@spaces}{      Package graphicx or graphics not loaded    }{See the gnuplot documentation for explanation.    }{The gnuplot epslatex terminal needs graphicx.sty or graphics.sty.}    \renewcommand\includegraphics[2][]{}  }  \providecommand\rotatebox[2]{#2}  \@ifundefined{ifGPcolor}{    \newif\ifGPcolor
    \GPcolortrue
  }{}  \@ifundefined{ifGPblacktext}{    \newif\ifGPblacktext
    \GPblacktexttrue
  }{}    \let\gplgaddtomacro\g@addto@macro
    \gdef\gplbacktext{}  \gdef\gplfronttext{}  \makeatother
  \ifGPblacktext
        \def\colorrgb#1{}    \def\colorgray#1{}  \else
        \ifGPcolor
      \def\colorrgb#1{\color[rgb]{#1}}      \def\colorgray#1{\color[gray]{#1}}      \expandafter\def\csname LTw\endcsname{\color{white}}      \expandafter\def\csname LTb\endcsname{\color{black}}      \expandafter\def\csname LTa\endcsname{\color{black}}      \expandafter\def\csname LT0\endcsname{\color[rgb]{1,0,0}}      \expandafter\def\csname LT1\endcsname{\color[rgb]{0,1,0}}      \expandafter\def\csname LT2\endcsname{\color[rgb]{0,0,1}}      \expandafter\def\csname LT3\endcsname{\color[rgb]{1,0,1}}      \expandafter\def\csname LT4\endcsname{\color[rgb]{0,1,1}}      \expandafter\def\csname LT5\endcsname{\color[rgb]{1,1,0}}      \expandafter\def\csname LT6\endcsname{\color[rgb]{0,0,0}}      \expandafter\def\csname LT7\endcsname{\color[rgb]{1,0.3,0}}      \expandafter\def\csname LT8\endcsname{\color[rgb]{0.5,0.5,0.5}}    \else
            \def\colorrgb#1{\color{black}}      \def\colorgray#1{\color[gray]{#1}}      \expandafter\def\csname LTw\endcsname{\color{white}}      \expandafter\def\csname LTb\endcsname{\color{black}}      \expandafter\def\csname LTa\endcsname{\color{black}}      \expandafter\def\csname LT0\endcsname{\color{black}}      \expandafter\def\csname LT1\endcsname{\color{black}}      \expandafter\def\csname LT2\endcsname{\color{black}}      \expandafter\def\csname LT3\endcsname{\color{black}}      \expandafter\def\csname LT4\endcsname{\color{black}}      \expandafter\def\csname LT5\endcsname{\color{black}}      \expandafter\def\csname LT6\endcsname{\color{black}}      \expandafter\def\csname LT7\endcsname{\color{black}}      \expandafter\def\csname LT8\endcsname{\color{black}}    \fi
  \fi
  \setlength{\unitlength}{0.0500bp}  \begin{picture}(5760.00,4032.00)    \gplgaddtomacro\gplbacktext{      \csname LTb\endcsname      \put(1513,704){\makebox(0,0)[r]{\strut{} 0}}      \put(1513,1183){\makebox(0,0)[r]{\strut{} 0.05}}      \put(1513,1661){\makebox(0,0)[r]{\strut{} 0.1}}      \put(1513,2140){\makebox(0,0)[r]{\strut{} 0.15}}      \put(1513,2618){\makebox(0,0)[r]{\strut{} 0.2}}      \put(1513,3097){\makebox(0,0)[r]{\strut{} 0.25}}      \put(1513,3576){\makebox(0,0)[r]{\strut{} 0.3}}      \put(1645,484){\makebox(0,0){\strut{} 0}}      \put(2258,484){\makebox(0,0){\strut{} 1}}      \put(2870,484){\makebox(0,0){\strut{} 2}}      \put(3483,484){\makebox(0,0){\strut{} 3}}      \put(4095,484){\makebox(0,0){\strut{} 4}}      \put(4708,484){\makebox(0,0){\strut{} 5}}      \put(3176,154){\makebox(0,0){\strut{}$\mu$ in bohr$^{-1}$}}      \put(971,2140){\rotatebox{90}{\makebox(0,0){\strut{}Excitation energies in hartree}}}    }    \gplgaddtomacro\gplfronttext{      \csname LTb\endcsname      \put(3388,2499){\makebox(0,0)[l]{\strut{}$1 ^1\Sigma_\text g^+ \to 1 ^3\Sigma_\text u^+ $}}      \csname LTb\endcsname      \put(3388,2235){\makebox(0,0)[l]{\strut{}$1 ^1\Sigma_\text g^+ \to 1 ^1\Sigma_\text u^+ $}}      \csname LTb\endcsname      \put(3388,1971){\makebox(0,0)[l]{\strut{}$1 ^1\Sigma_\text g^+ \to 2 ^1\Sigma_\text g^+ $}}    }    \gplbacktext
    \put(0,0){\includegraphics{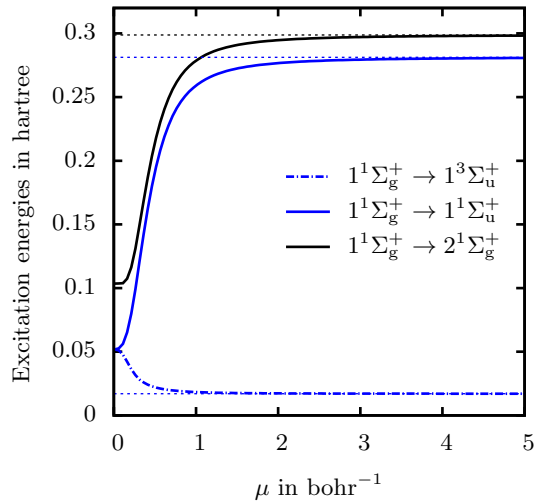}}    \gplfronttext
  \end{picture}\endgroup
  \caption{Excitation energies $\Delta {\cal E}_k^{\mu} = {\cal E}_k^{\mu} - {\cal E}_0^{\mu}$ (in hartree) of the H$_2$ molecule at 3 times the equilibrium internuclear distance as a function of $\mu$ (in bohr$^{-1}$). The excitation energies of the physical system $\Delta E_k = \Delta {\cal E}_k^{\mu\to\infty}$ are plotted as horizontal dotted lines.
    \label{fig:h2_eq3_dacpvtz_exc_0}
  }
\end{figure}

Finally, we consider the interesting case of the dissociation of the H$_2$ molecule. 
The first excitation energies at three times the equilibrium distance are shown in Figure~\ref{fig:h2_eq3_dacpvtz_exc_0}. 
With increasing bond distance, the $1\sigma_\text g$ and $1\sigma_\text u$ molecular orbitals become degenerate. Consequently,
the KS excitation energy for the single excitation $1\sigma_\text g \to 1\sigma_\text u$ goes to zero. 
Moreover, the KS excitation energy for the double excitation $(1\sigma_\text g)^2 \to (1\sigma_\text u)^2$ also goes to zero (albeit more slowly).
This behaviour is in contrast to that of the physical system, where 
only the excitation energy to the triplet $1^3\Sigma_\text u^+$ state goes to zero, 
whilst those to the singlet $1^1\Sigma_\text u^+$ state and the $2^1\Sigma_\text g^+$ state 
(the latter connected to the double excitation in the KS system) go to finite values. 

Clearly, the excitation energies of KS theory are poor approximations to the exact excitation energies, making it difficult to recover 
from these poor starting values in practical linear-response TDDFT calculations. 
As $\mu$ increases from $\mu=0$, the excitation energies to the singlet $1^1\Sigma_\text u^+$ and $2^1\Sigma_\text g^+$ states vary abruptly, 
rapidly approaching the physical values. 
This sensitivity to the inclusion of the electron--electron interaction is a clear signature of strong static correlation effects, 
emphasizing the importance of a multi-determinantal description in such situations. 
At $\mu \approx 0.4-0.5$, 
the $1^1\Sigma_\text u^+$ and $2^1\Sigma_\text g^+$ excitation energies, 
although still too low, are much better approximations than the KS excitation energies, 
constituting a strong motivation for range-separated multi-determinantal approaches in linear-response theory.

\section{Conclusions}
\label{sec:conclusion}

We have studied the variation of total energies and excitation energies along a range-separated adiabatic connection, 
linking the non-interacting KS system ($\mu=0$) to the physical system ($\mu\to\infty$) by progressively switching 
on the long-range part of the electron--electron interaction with the range-separation parameter $\mu$, 
whilst keeping the ground-state density constant. This behaviour is
of interest for the development and analysis of range-separated DFT schemes for the calculation of excitation energies, 
such as the linear-response range-separated schemes of Refs.\;\onlinecite{RebSavTou-MP-13,FroKneJen-JCP-13,HedHeiKneFroJen-JCP-13}.

Reference calculations were performed for the He and Be atoms and the H$_2$ molecule. Except when $\mu$ is large, 
the ground- and excited-state total energies along the adiabatic connection are poor approximations to the corresponding energies of the physical system. 
On the other hand, the excitation energies are good approximations to the excitation energies of the physical system for most of the adiabatic 
connection curve, except close to the KS system ($\mu=0$). In particular, the excitation energies obtained at $\mu \approx 0.4-0.5$, 
typically used in range-separated DFT calculations, are significantly better approximations to the exact excitation energies than 
are the KS excitation energies. 
This behaviour appears to be particularly evident for situations involving strong static correlation effects and double excitations, 
as observed for the dissociating H$_2$ molecule. 

These observations suggest that the excitation energies of the long-range interacting Hamiltonian in range-separated DFT may be useful 
as first estimates of the excitation energies of the physical system. However, if one cannot
afford to use large $\mu$ values ($\mu > 2-3$), these excitation energies should be considered only as starting approximations, 
suitable for correction by, for example, linear-response range-separated theory.

In future work, we will utilize the present reference data to assess the approximations made in practical linear-response range-separated 
schemes, where the long-range contribution is treated, for example, at the Hartree--Fock, MCSCF or SOPPA levels of theory, 
while the short-range part is described by semi-local density-functional approximations. We will also use the results of this work 
to guide the development of time-independent range-separated DFT methods for the calculation of excitation energies as alternatives 
to linear-response schemes---in particular, for  methods based on perturbation theories~\cite{Gor-PRA-96,Filippi1997} or extrapolations~\cite{Sav-JCP-11,Sav-JCP-14} along the adiabatic connection.

\begin{acknowledgments}
E.~R. and J.~T. gratefully acknowledge the hospitality of the Centre for Theoretical and Computational Chemistry (CTCC),
University of Oslo, where part of this research was done. E.~R. also thanks A. Borgoo and S. Kvaal for helpful discussions.
T.~H.~acknowledges support from the Norwegian Research Council through the
CoE Centre for Theoretical and Computational Chemistry (CTCC) Grant
No.\ 179568/V30 and the Grant No.\ 171185/V30 and through the European
Research Council under the European Union Seventh Framework Program
through the Advanced Grant ABACUS, ERC Grant Agreement No.\ 267683.
A.~M.~T.~is grateful for support from the Royal Society University Research Fellowship scheme.
\end{acknowledgments}

\end{document}